\documentclass{jpp}
\usepackage{graphicx}

\graphicspath{{./}{figures/}}

\usepackage[utf8]{inputenc}
\usepackage[T1]{fontenc}
\usepackage{amsmath}
\usepackage{amssymb}
\usepackage{dsfont}

\usepackage{unicode}

\newcommand{\vpar}{\ensuremath{v_{\parallel}}}
\newcommand{\vperp}{\ensuremath{v_{\perp}}}

\providecommand\bnabla{\boldsymbol{\nabla}}
\providecommand\bkappa{\boldsymbol{\kappa}}

\providecommand\bnabla{\boldsymbol{\nabla}}

\def \al {\mbox{$\alpha$}}

\def \kperp {\mbox{$k_{\perp}$}}
\def \Lpar {\mbox{$L_{\parallel}$}}
\def \vth {\mbox{$v_{\mathrm{T}}$}}

\def \epseff {\mbox{$\epsilon_\text{eff}$}}

\def \l {\mbox{$\ell$}}
\def \lp {\mbox{$\ell^{'}$}}

\def \eps {\mbox{$\epsilon$}}

\renewcommand \o {\mbox{$\omega$}}

\def \os {\mbox{$\omega_*$}}

\def \ost {\mbox{$\widetilde{\omega}_*$}}

\def \odt {\mbox{$\widetilde{\omega}_d$}}

\def \Reff {\mbox{$R_{\mathrm{eff}}$}}

\usepackage{xcolor}
\usepackage{pagecolor}
\usepackage{lipsum}  

\shorttitle{CG for QI}
\shortauthor{G. T. Roberg-Clark, P. Xanthopoulos, G. G. Plunk, S. Stroteich}

\title{Critical gradient optimization for quasi-isodynamic stellarators}

\author{G. T. Roberg-Clark\aff{1}
  \corresp{\email{gar@ipp.mpg.de}},
  P. Xanthopoulos \aff{1},
  G. G. Plunk \aff{1},
  S. Stroteich \aff{2}
  }

\affiliation{\aff{1}Max-Planck-Institut F\"ur Plasmaphysik, D-17491, Greifswald, Germany \aff{2}Institute of Physics, University of Greifswald, 17489 Greifswald, Germany }

\begin{document}

\maketitle

\begin{abstract}

We present new and updated methods for reducing transport caused by electrostatic ion temperature gradient (ITG) driven turbulence in quasi-isodynamic (QI) configurations. We first show an updated model for the threshold (critical) gradient of localized toroidal ITG modes. It is then argued that it is desirable for ITG modes to ``split'' and localize in separate curvature drift wells, which is leveraged to produce a six-field-period QI configuration with a high critical gradient. We show that the destabilizing effect of kinetic electrons \citep{Costello2025} on localized ITG modes can be minimized in a magnetic field structure, which we dub “inverse mirror”. Applying a general optimization target that improves ITG stability above the critical gradient yields an inverse mirror configuration which maintains heat fluxes below or equal to the W7-X high mirror configuration for a range of applied density gradients.

\end{abstract}

\section{Introduction} \label{sec:intro}

A central issue in magnetic confinement fusion is the reduction of heat and particle transport to the edge of the fusion device such that temperature and density profiles remain conducive to fusion. While confinement can be increased by scaling up the size of a fusion device, it may be more practical in the end to design smaller devices with better confinement properties that are cheaper to build. In this vein it makes sense to reduce transport resulting from small-scale turbulence as long as it results in steeper temperature profiles. Contrary to what one might think, this does not mean that the possibly favorable effects of turbulence (e.g. exhaust of impurity particles \citep{Garcia-Regana2021a}) are to be removed from the problem. Instead, we can imagine increasing the temperature gradient for which the same amount of transport is produced, increasing fusion power in the core of the device while maintaining power and exhaust balance. Here we continue a line of work exploring how to model and increase the threshold, the critical gradient or CG, for ion-temperature-gradient modes, to potentially increase temperature gradients in stellarator device designs. We focus on stellarators but the ideas can be applied to tokamaks as well, since the quantities used in the model are taken from local magnetic field line geometry in toroidal systems.

Previous work \citep{Roberg-Clark2021,Roberg-Clark2022} modeled the minimum CG of the system - the gradient at which linear ITG modes are excited - as arising from either global magnetic shear or a phase mixing effect produced by unfavorable (called ``bad'') magnetic field line curvature \citep{Baumgaertel2013}, in line with previous work in tokamak research \citep{Romanelli1989,Jenko2001}. While these results gave basic insight into the physics of the CG in 3D toroidal geometry, they suggested optimization strategies that led to high magnetic shear and MHD unstable configurations that were not conducive to low overall transport.

In a subsequent paper the focus was shifted away from the onset of any linear modes \citep{Zocco2018} to that of linear modes that are strongly localized to regions of bad curvature and able to overcome parallel Landau damping \citep{Roberg-Clark2023,Roberg-Clark2024}. These modes are thought to be the most detrimental for transport, with relatively high growth rates indicative of strong curvature drive. The latter is typically correlated with larger perpendicular wavenumbers, since the binormal wavenumber $k_{\alpha}$ is multiplied with the geometric curvature factor in the gyrokinetic equation. Here $\alpha$ is defined through the magnetic field representation in field following (Clebsch) representation, $\mathbf{B}=\bnabla \psi \times \bnabla \al$, where $\psi$ is a flux surface label and $\alpha$ labels the magnetic field line on the surface. The perpendicular wave vector is expressed as $\mathbf{k_{\perp}} = k_{\alpha} \bnabla \alpha + k_{\psi} \bnabla \psi$, where $k_{\alpha}$ and $k_{\psi}$ are constants, so the variation of $\mathbf{k_{\perp}}(\ell)$ stems from that of the geometric quantities $\bnabla \alpha$ and $\bnabla \psi$, with $\l$ the field-line-following (arc length) coordinate. Even though mixing length estimates ($Q \propto \gamma / \kperp^{2}$) suggest high wavenumber modes will contribute relatively less to overall transport, it has nonetheless been found that targeting these modes, via increasing $\bnabla \alpha$, leads to noticeably smaller values of heat flux in the nonlinear system. 

The strategy of enhancing $\bnabla \alpha$ also complements work showing that reducing the gradient of the radial coordinate $\bnabla \psi$ in the region of bad curvature often suppresses nonlinear heat fluxes \citep{Angelino2009,Xanthopoulos2014a,Goodman2024,Landreman2025}, since for a fixed magnetic field $B$, reducing $\bnabla \psi$ will increase $\bnabla \alpha$ \citep{Roberg-Clark2024}. One way to view this strategy is that the gyro-Bohm scaling for heat fluxes permits a rescaling of the temperature gradient by the factor of $\bnabla \psi$ \citep{Stroteich2022}. This rescaling factor is related to the fact that in real space coordinates any increase in the gradient of the radial coordinate leads to a corresponding increase in the physical temperature gradient. That said, we still emphasize that $\bnabla \alpha$ not only contains this effective gradient (via inverse proportionality with $\bnabla \psi$ at constant $B$) but also magnetic shear effects, not included in $\bnabla \psi$, that can strongly damp ITG modes. Using insights from these results we start to incorporate physics of ITG modes above the CG in results shown at the end of the paper.

We apply the methods developed here to a class of omnigenous magnetic fields called quasi-isodynamic (QI) stellarators. Recent results \citep{Goodman2024,Plunk2024} have shown that QI magnetic fields have favorable properties related to turbulence stemming from low zonal flow damping and that this can be combined with a reduction of $\bnabla \psi$ to significantly reduce ITG turbulence. In this paper we show how to generalize a previous model so that the CG method can be utilized for QI stellarators and any other toroidal confinement magnetic field of interest.

The paper proceeds as follows. In section \ref{sec:ITG} we present the linear gyrokinetic equation and discuss the updated CG model. In section 3 we briefly cover how omnigenity is targeted for QI optimization, and in section 4 we present two optimized configurations that illustrate how the models can be successfully leveraged to improve ITG stability. Readers most interested in the optimization results rather than gyrokinetic modelling can skip to this section. We conclude the paper in section \ref{sec:conc}.

\section{ITG physics}

\subsection{Linear gyrokinetic equation} \label{sec:ITG}

We use the standard gyrokinetic system of equations \citep{Brizard2007} to describe electrostatic fluctuations destabilized along a thin flux tube tracing a magnetic field line. The ballooning transform \citep{Dewar1983a,Connor1978} is used to separate out the fast perpendicular (to the magnetic field) scale from the slow parallel scale. We assume Boltzmann-distributed (adiabatic) electrons, thus solving for the perturbed ion distribution $g_{i}(\vpar,\vperp,\l,t)$, defined to be the non-adiabatic part of $\delta f_{i}$ ($\delta f_{i}=f_{i}-f_{i0})$ with $f_{i}$ the ion distribution function and $f_{i0}$ a Maxwellian. The electrostatic potential is $\phi(\mathbf{\l})$, and $\vpar$ and $\vperp$ are the particle velocities parallel and perpendicular to the magnetic field, respectively. The gyrokinetic equation reads
\begin{equation}
i\vpar \frac{\partial g}{\partial \ell} + (\omega - \odt)g = \varphi J_0(\omega - \ost^{T})f_0\label{gk-eqn}
\end{equation}
where $\omega$ is the mode frequency, $\ost^{T} = (Tk_{\alpha}/q)\mathrm{d}\ln T/\mathrm{d}\psi \left(v^2/\vth^2 - 3/2\right)$ is the diamagnetic frequency, and $J_{0} = J_{0}(k_{\perp}(\l)v_{\perp}/\Omega(\l))$ is the Bessel function of zeroth order. The thermal velocity is $\vth = \sqrt{2T/m}$, the thermal ion Larmor radius is $\rho = \vth/(\Omega\sqrt{2})$, $n$ and $T$ are the background ion density and temperature, $q$ is the ion charge, $\varphi = q\phi/T$ is the normalized electrostatic potential, and $\Omega=q B/m$ is the cyclotron frequency, with $B=|\mathbf{B}|$. The magnetic drift frequency is $\odt = (1/\Omega)\mathbf{k_{\perp}} \cdot \left[ ( \mathbf{b} \times \bkappa)\vpar^2 + (1/B)(\mathbf{b} \times \bnabla B)  \vperp^2/2 \right]$, with $\bkappa = \mathbf{b} \cdot\bnabla\ \mathbf{b}$ and $\mathbf{b}=\mathbf{B}/B$. For simplicity in this analysis, we set $k_{\psi}=0$. We hereafter focus on the curvature drift, $K_{d}(\ell) \equiv a^{2}{\bnabla}\alpha \cdot \mathbf{b} \times \bkappa$, referring to $K_{d}(\ell)$ as the ``drift curvature'' and to individual regions of bad curvature along the field line (where $K_{d}>0$) as drift wells. The curvature drift is used since, at finite plasma $\beta$, it experiences a shift relative to the grad-B drift proportional to $\bnabla P/B^{2}$, with $P$ the plasma pressure and the sign of this term can make bad curvature worse. We define a radial coordinate $r=a\sqrt{\psi/\psi_{edge}}$, with $a$ the minor radius corresponding to the flux surface at the edge, and $\psi_{edge}$ the toroidal flux at that location. The temperature gradient scale length is measured relative to the minor radius, $a/L_{T}=-(a/T)\mathrm{d}T/\mathrm{d}r$.

Finally, the gyrokinetic system is completed by quasineutrality,
\begin{equation}
\int d^3{\bf v} J_{0} g = n(1 + \tau) \varphi\label{eqn:qn},
\end{equation}
with $\tau=|q_{e}|T/(qT_{e})$, $T_{e}$ the electron temperature, and $q_{e}$ the electron charge.

\subsection{Previous CG model}

To start the discussion of CG models we quote an earlier expression from \citep{Roberg-Clark2023} that incorporates $\bnabla \alpha$ as well as the parallel connection length $\Lpar$ of the geometry. This expression was obtained by fitting coefficients to a numerical solution of the linear dispersion relation for localized toroidal modes and reads

\begin{equation}
\frac{a}{L_{T,crit}} = \left( \frac{1+\tau}{2} \right) \frac{a}{\Reff} \begin{cases} 
      2.58 + 4.926 \left( \pi a|\bnabla \alpha| \frac{\Reff}{\Lpar} \right)^{2}, & \left(\pi a|\bnabla \alpha| \frac{\Reff}{\Lpar} \right) < 0.78 \\
      7.5 \: \pi a|\bnabla \alpha| \frac{\Reff}{\Lpar}, & \left(\pi a|\bnabla \alpha| \frac{\Reff}{\Lpar} \right) \geq 0.78.
   \end{cases} \label{eqn:oldcg}
\end{equation}

Here $\Reff$ in this case was determined by doing a local quadratic fit to the profile of $K_{d}$ and selecting the peak value of the local fit $(1/\Reff=K_{d,\text{peak}})$, $\bnabla \alpha$ was chosen at the location of the peak fitted $K_{d}$, and $\Lpar$, the parallel connection length, was the distance between two points where the sign of $K_{d}$ flipped, defining the bad curvature region for the fitting. The quadratic fit was used as a form of coarse-graining - an averaging procedure meant to draw out the underlying behavior of a smooth background in the presence of oscillations. We found that generally, for stellarators with a large number of field periods and a vacuum magnetic well, the short connection length (or weak curvature) limit $\pi a|\bnabla \alpha| \Reff/\Lpar \geq 0.78$ tends to be the most relevant and we therefore focus on this limit in what follows.

\begin{figure}
    \centering
    \includegraphics[width=0.50\linewidth]{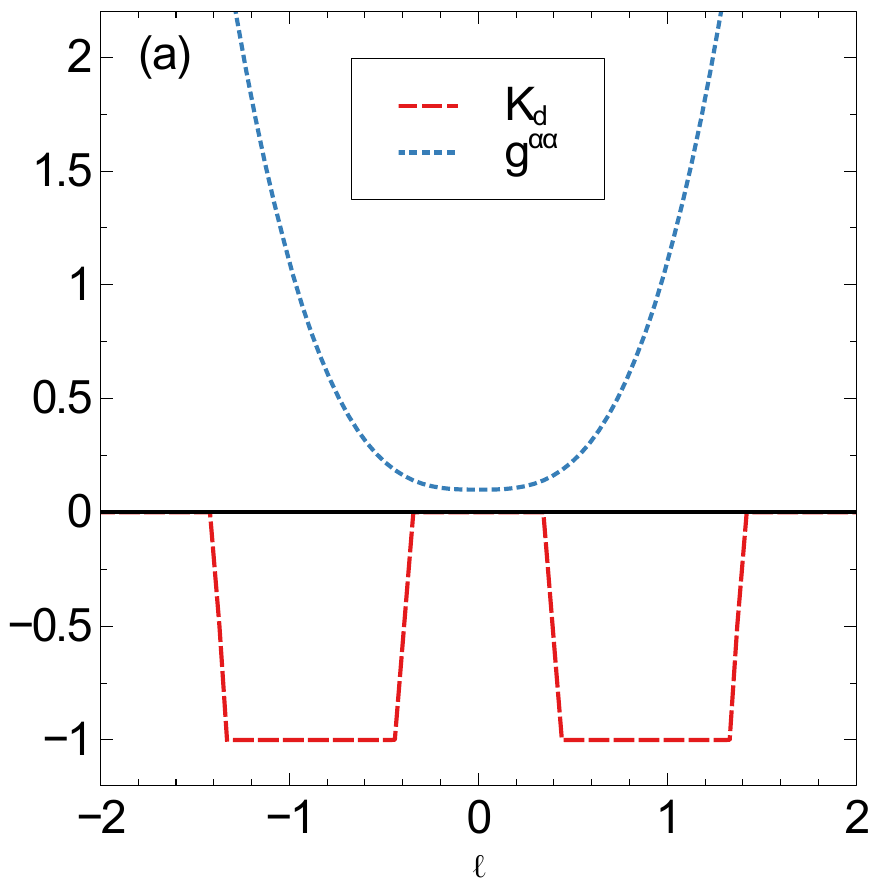}
    \includegraphics[width=0.40\linewidth]{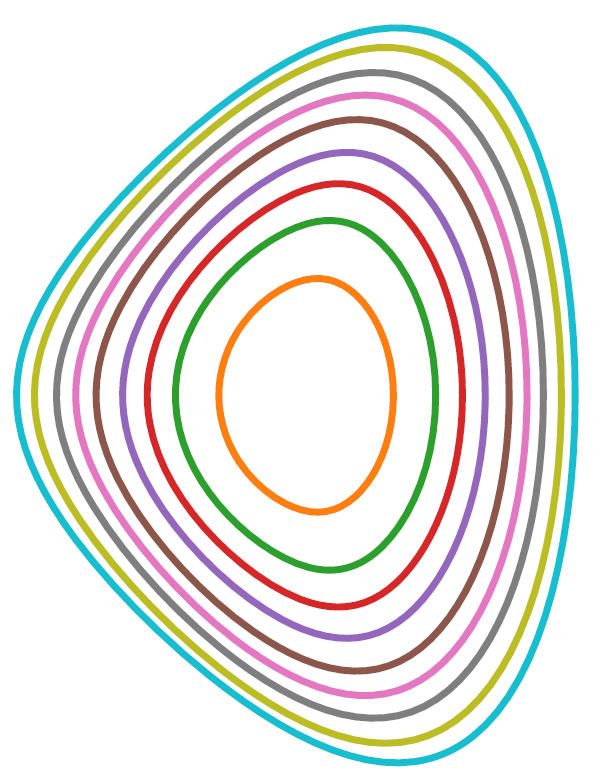}
    \caption{Left: Simplified toy model geometry for the drift curvature $K_{d}$ and squared gradient of the binormal coordinate ($g^{\alpha\alpha} = |\bnabla \alpha|^{2}$) depicted along a magnetic field line ($\l$ coordinate), with split drift curvature wells and secularly increasing $g^{\alpha\alpha}$. A QI stellarator, or reverse triangularity tokamak with large magnetic shear and flat surfaces near the outboard midplane could qualitatively achieve such geometry profiles. Right: magnetic flux surfaces and outer boundary for a QI configuration at zero toroidal VMEC angle with a ``reverse-D'' shape at the maximum of $B$. Note the relative compression of the surfaces on the outboard. Roughly constant magnetic field (as the poloidal angle is traversed) is achieved by offsetting the $1/r$ scaling of $B$ with surface expansion on the inboard and compression on the outboard.}
    \label{fig:qinegTok}
\end{figure}

A basic issue with the previous method is that drift curvature wells in general do not take the form of a low order polynomial function. It is often the case that the drift curvature at the outboard midplane, where ITG turbulence often peaks in gyrokinetic simulations, has a peculiar form for QI configurations. Instead of a single drift well that can be approximated by a sine function or a quadratic curve, a typical quasi-isodynamic configuration instead shows a ``split'', symmetric about the point mentioned above, into two distinct drift wells with a null at the central location. The null point is a consequence of having straight magnetic field lines in real space at the maximum of the magnetic field strength $B$. While the previous method can perhaps accurately model $g^{\alpha \alpha}$ with its value at the center of the drift well for quasisymmetric configurations, $g^{\alpha \alpha}$ can in fact vary significantly. \cite{Landreman2025} found that $g^{\psi \psi}$, when averaged over the drift well, was better correlated with heat fluxes than the value chosen at a single point, in this case the outboard midplane location where turbulence often causes the most transport. These results lend credence to the idea that drift-well-averaged quantities are the most relevant to transport and should lead to improvements over the old method.

We note that the approximate geometric situation described above is not limited to QI stellarators, since tokamaks with reverse triangularity often exhibit similar double-well behavior, see Fig. 15 (b) and (c) of \cite{Balestri2024}. We show a toy model geometry in Fig. \ref{fig:qinegTok} which has qualitative elements that can arise in both geometries, in that the drift curvature contains a null (or nearly zero) region of finite width and $g^{\alpha \alpha} = |\bnabla\alpha|^2$ is amplified over the region of bad curvature as a result of magnetic shear. Negative triangularity tokamaks have apparent ITG stability benefits such as relatively high CGs \citep{Balestri2024}, an effect which stellarators may emulate, despite belonging to a different geometry class. A CG model capable of handling such cases would thus also allow modelling of more complex stellarator and tokamak cases while avoiding possible issues related to evaluating $g^{\alpha \alpha}$ at a single point. Furthermore, such a model could be used in optimization to force the parallel connection length of localized toroidal modes to lie within the now-separated drift wells, effectively doubling the CG (by halving $\Lpar$ in expression (\ref{eqn:oldcg}). We refer to this as ``splitting the mode'' (section 2.4) and now describe how we refine the CG model to handle such cases.

\subsection{Updated CG model}

We now imagine coarse-graining of the geometry as a kind of field line average that takes place over a single region of bad curvature in which the sign of the bad curvature does not reverse. We update the previous model (Roberg-Clark et al. 2023) by defining the RMS average, 
\begin{equation} \label{eqn:galphaCG}
\langle (\cdot \cdot\cdot) \rangle_{CG} (\l) = \left[ \frac{ \ \int\Theta(K_{d}(\lp)) (\cdot\cdot\cdot)^{2}(\lp) \Theta_{w}(\l,\lp) d \lp }{  \int \Theta(K_{d}(\lp)) \Theta_{w}(\l,\lp) d \lp} \right]^{1/2},
\end{equation}
to calculate quantities like the average of $|\bnabla \alpha|$ instead of choosing the central peak fitted value of $|\bnabla \alpha|$ as before. The ``observation point'' $\l$ effectively sets the mode location, as enforced by the Heaviside function $\Theta_{w}(\l,\lp)$, which is $1$ if $K_{d}$ has no sign changes between $\l$ and $\lp$ and zero otherwise. This restricts the integration to single drift wells, where the other Heaviside function, $\Theta(K_{d}(\lp))$, is $1$ where $K_{d}$ is negative (destabilizing) and zero otherwise. The parallel connection length is calculated by finding the width of this region, namely
\begin{equation}
    \Lpar(\l)_{CG} = \int \Theta(K_{d}(\lp)) \Theta_{w}(\l,\lp) d \lp,
\end{equation}
while the average curvature drive at marginal stability is defined as
\begin{equation}
    \left( \frac{a}{\Reff} \right)_{CG}(\l) = \frac{1}{\Lpar_{CG}(\l)} \int K_{d}(\lp) \Theta(K_{d}(\lp)) \Theta_{w}(\l,\lp) d \lp.
\end{equation}
Note that in contrast to the previous model, the integration is carried out such that the endpoints (regions where the sign of $K_{d}$ flips between two grid points) are included with linear interpolation in the integral. Armed with more general methods for calculating averages, we express the new CG along a single field line $\alpha$ as
\begin{equation}    
\frac{a}{L_{T,crit}}(\alpha) = \text{Min}_{\l} \left[ \pi \left( \frac{1+\tau}{2} \right)
      \left (3.75 \: \langle a|\bnabla \alpha| \rangle_{CG}(\l) \frac{a}{\Lpar_{CG}(\l)} \right) \right], \label{eqn:CGopt}
\end{equation}
which ignores any explicit dependence on $\Reff$ (consistent with the weak curvature limit) and uses a coefficient of half the size compared to the old model for the $\Lpar$ term. Implemented as an optimization target, the method scans over all observation points $\l$ present in the given flux tube and finds the location corresponding to the minimum CG. The lack of explicit dependence on $\Reff$ means that during optimization the size of curvature will be determined by properties other than the CG, like the aspect ratio, number of field periods, imposition of a vacuum magnetic well, and addition of plasma $\beta$, which modifies the drift curvature profile. In this finite $\beta$ limit we use the curvature drift, as defined through $K_{d}$ in section 2.1.
The target is then defined as
\begin{equation}    
f_{CG}(r)=\left(\text{Min}_{\alpha}\left[\frac{a}{L_{T,crit}}(\alpha)\right]-2.0\right)^{2} \label{eqn:CGf}
\end{equation}
which can be used to improve the CG during optimization on a specific surface $r$. For all of the cases in this paper, we optimize at $r/a=0.5$.

\subsection{Splitting the mode}

\begin{figure}
    \centering
    \includegraphics[width=0.99\linewidth]{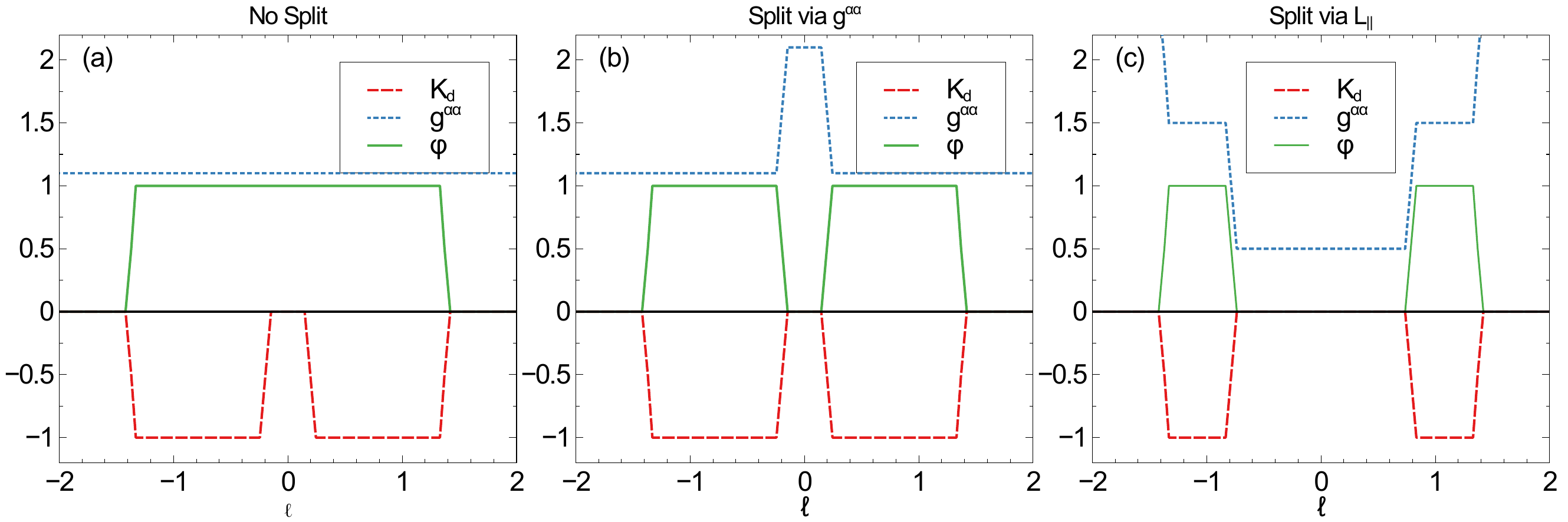}
    \caption{\textit{Splitting ITG modes} -- Idealized field line geometry metrics at the typically most unstable location on the outboard midplane of a QI magnetic field configuration.(a) No attempt is made to split the mode geometrically across the standard gap at $\ell=0$. Thus the mode can average the drift curvature across the gap. We surmise that $(a/\Reff)_{CG}$ will be similar to a case with no gap. The mode freely ``tunnels'' across the gap to encompass an $\Lpar$ of the entire well (achieving a low critical gradient) with similar curvature drive to modes that localize on each side of the gap. (b) A barrier is imposed by the geometry through $g^{\alpha \alpha} = |\bnabla\alpha|^2$, reducing the growth rate of the mode that tunnels across the gap. (c) Separating the two wells reduces the average curvature drive significantly for the tunneling mode, as similarly portrayed in Fig. \ref{fig:qinegTok}. Additional shear amplification of $g^{\alpha \alpha}$ further stabilizes the localized modes.}
    \label{fig:split}
\end{figure}

Having found a critical gradient formula that is compatible with well-splitting, we can discuss how this effect is realized in practice.  In particular, we imagine two different ways to ensure that the parallel connection length of toroidal ITG modes is halved by splitting the mode. The goal here, via manipulating the field line geometry, is to decouple the behavior of modes that straddle both wells and those that localize to individual wells. The localized modes should then be further stabilized by the geometry. 

To illustrate the common (un-optimized) situation for QI stellarators, we first show a case where no attempt is made to split the mode, aside from having a zero point of curvature [Fig. \ref{fig:split}(a)]. In this case the mode can simply ``tunnel'' across the gap at little cost to the average value of the curvature (the gap is small compared to the overall width of the region). The end result is a mode with a parallel connection length containing both drift wells that is not significantly affected, in terms of curvature drive or parallel dynamics, by a null point in the curvature.

A way to more effectively split the mode was already shown in Fig. \ref{fig:qinegTok} -- to make the gap between the two resulting wells large, such that the gap is comparable to or larger than the width of the two wells, as also shown in Fig. \ref{fig:split}(c). The modes which are strongly driven by curvature will be forced to localize in each separate well, and their stability will then be set by the local drift well behavior, while the mode that spreads across the barrier will experience a reduced curvature drive $(a/\Reff)_{CG}$ as long as the curvature does not increase in proportion to the widening of the gap. Note this means that residual turbulence may grow at a CG below that of the more localized modes, since they have a larger $\Lpar$, which we believe will produce a ``foot'' in plots of nonlinear heat flux versus temperature gradient \citep{Zocco2018,Zocco2022}. A second way to split the well is to insert a barrier through which ITG modes with strong curvature drive cannot tunnel, in this case by increasing $\bnabla \alpha$ at the central location $\l=0$ on the flux tube \citep{Roberg-Clark2021} [Fig. \ref{fig:split}(b)]. The mode that tunnels suffers reduced growth rates as a result. We show stellarator configurations which utilize the first method later in the paper.

As already mentioned, the mode splitting (starting with a null point in the drift curvature), if it is present or nearly occurring in a vacuum magnetic equilibrium, can be altered by the shift in the curvature drift in the gyrokinetic equation brought about by a finite $\beta$ term proportional to $\bnabla p / B^{2}$. This term can make bad curvature worse since on the outboard the temperature gradient and normal curvature vectors tend to be parallel. As such, the curvature drift may capture a lower CG from an increase in $\Lpar_{CG}$ in finite $\beta$ systems.

\subsection{ITG instability above the threshold} \label{sec:stiff}
Once the ITG modes are localized to individual drift wells, it remains to be seen how stable they will be. Growth rates for these modes depend on a variety of geometric factors, such as the magnitude of $K_{d}$, $|\bnabla \alpha|$ and $|\bnabla \psi|$. Since, above the threshold, modes continue to localize in regions of bad curvature, we slightly modify the average in (\ref{eqn:galphaCG}) by weighting it with the magnitude of curvature:

\begin{equation} \label{eqn:galpha}
\langle (\cdot\cdot\cdot) \rangle_d (\l) = \left[ \frac{ \int \: K_{d}(\lp) \Theta(K_{d}(\lp)) (\cdot\cdot\cdot)^{2}(\lp) \Theta_{w}(\l,\lp)d \lp}{\int K_{d}(\lp) \Theta(K_{d}(\lp))\Theta_{w}(\l,\lp) d \lp} \right]^{1/2} .
\end{equation}
This factor will take into account surface compression (through relating $\bnabla \psi$ to $\bnabla \alpha$ through $\mathbf{B}$) and local and global shear effects that stabilize the localized ITG modes above the CG.

\subsection{Kinetic electrons} \label{sec:kinetic}

\begin{figure}
    \centering
    \includegraphics[width=0.99\linewidth]{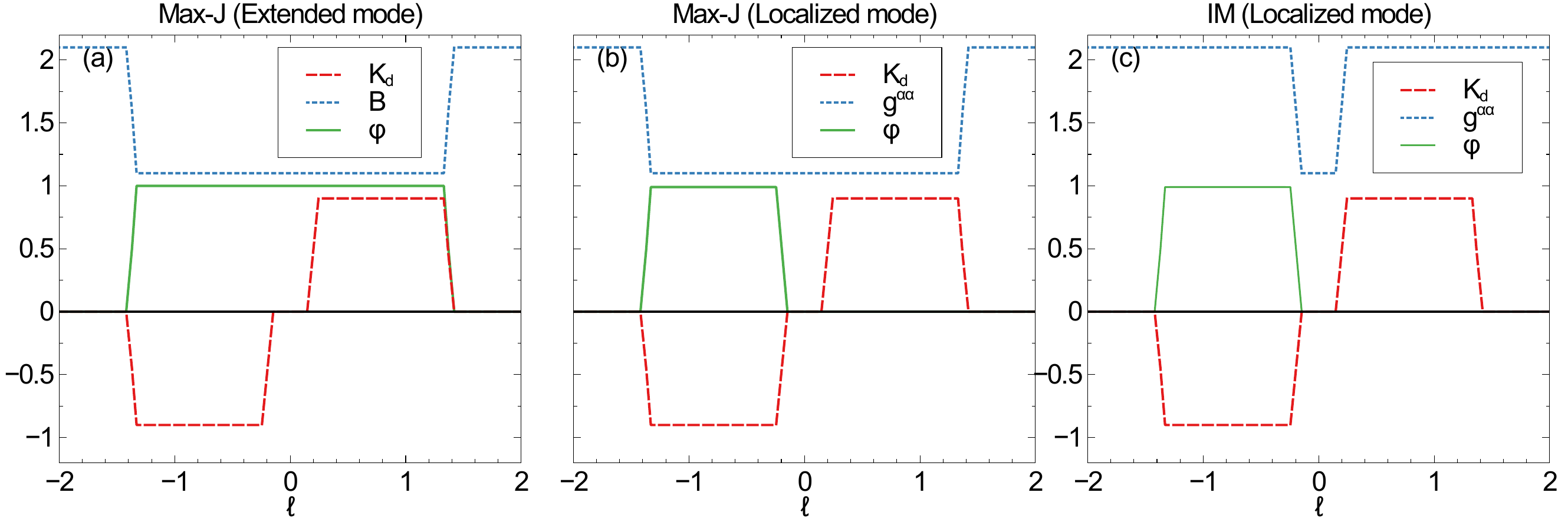}
    \caption{\textit{IM configurations sidestep mode inertia} -- Idealized schematics for different QI magnetic field shapes along a magnetic field line with different, roughly constant ITG mode potentials but the same QI-like curvature profile $K_{d}$. (a) A standard broad minimum and narrow maximum which leverages the max-J property. The spread-out mode potential achieves no net curvature drive and is stable. (b) The same geometry but with sufficient drive such that the critical gradient for a localized ITG mode is exceeded, which sits only in the bad curvature well. The mode inertia effect now increases drive for the ITG mode roughly by the factor $\left(1-\sqrt{1-B_{\text{min}}/B_{\text{max}}} \right)^{-1}$. (c) An ``inverse mirror'' (IM) magnetic field with a narrow minimum and broad maximum avoids the mode inertia effect by removing trapped particles from the bad curvature region ($\langle \epsilon(\l) \rangle \rightarrow 0$).}
    \label{fig:IMB}
\end{figure}

An additional factor is the effect of kinetic electrons on ITG mode growth rates. The non-adiabatic response of electrons can significantly increase growth rates and nonlinear heat fluxes of ITG modes compared to the adiabatic electron case (see e.g. \cite{Proll2022}). Several strategies to combat this enhanced drive from trapped particles have been developed in recent years with regards to trapped electron modes, including calculations of the available energy for turbulence \citep{Mackenbach2022,Mackenbach2023,Duff2025}, modifying flux surface shaping in both tokamaks \citep{Garbet2024} and stellarators \citep{Gerard2024}, or directly targeting the overlap of trapped particles with drift curvature \citep{Proll2016,Gerard2023}. Recent theoretical work \citep{Costello2025} focused on the modification of linear, toroidal ITG mode growth rates,
\begin{equation}
\omega = \pm \sqrt{-\os_{i}\int_{-\infty}^{\infty}\frac{d\ell}{B} \hat{\o}_{di}(\ell)|\delta \phi|^{2} }
\left( \tau \int_{-\infty}^{+\infty}|\delta \phi |^{2} \frac{dl}{B} - \frac{\tau}{2} \Sigma_{j} \int_{\frac{1}{B_{min}}}^{\frac{1}{B_{max}}} \tau_{B,j} |\overline{\delta \phi_{j}}|^{2} d\lambda \right)^{-\frac{1}{2}}, \label{eqn:tdisp}
\end{equation}
where the bounce average is
\begin{equation}
  \overline{(\cdot\cdot\cdot)} = \frac{1}{\tau_{b}}\int^{\l_{2}}_{\l_{1}} \frac{(\cdot\cdot\cdot)}{\sqrt{1-\lambda B}} d\l
\end{equation}
and the bounce time is $\tau_{B} = \int^{\l_{2}}_{\l_{1}} (1-\lambda B)^{-\frac{1}{2}} d\l$. Several effects are at play in equation (\ref{eqn:tdisp}). The first is the integral along the field line of the drift curvature $\hat{\o}_{di}(\ell) \propto -K_{d}(\l)$, indicating stability based on the size and sign of the average curvature. 
Negative curvature in this notation indicates stability of the mode, and is a motivating factor for the max-J property \citep{Proll2012,Rodriguez2024a}, which has led to successful QI designs with reduced turbulence \citep{Goodman2024,Regana2025}. ITG modes can be completely stabilized in cases where the mode extent includes both regions of good and bad curvature, since the average over this region can be of the ``good'' sign [Fig. \ref{fig:IMB}(a)]. However, toroidal magnetic fields must have some regions of bad curvature, and we are exploring ITG modes that only exist in regions of bad curvature. The max-J property itself cannot stabilize such modes, since $|\delta \phi| =0$ in regions of good curvature, by definition [see Fig. \ref{fig:IMB}(b)]. Thus the average drive of these modes will be unaffected by the max-J property once they are driven unstable above the CG.

To reduce the growth rates of localized ITG modes in the kinetic electron regime, we are left with the options of reducing the magnitude of the bad curvature in the numerator or reducing the effect of non-adiabatic electrons in the denominator. As already stated, we would prefer to leave the former quantity alone during optimization. The latter effect was referred to as ``mode inertia'' \citep{Costello2025} and results from the presence of trapped particles capable of strongly destabilizing ITG modes. Note that in the limit of a square well magnetic field, for which $B=B_{min}$ in some central region bounded by step functions with height $B=B_{max}$, the second term in the denominator includes the trapped particle fraction $\epsilon = \sqrt{1-B_{min}/B_{max}}$. Since the overall denominator is positive-definite, it reduces in magnitude from the non-adiabatic response and leads to increased growth rates. To make an estimate for spatially varying magnetic fields with ITG modes localized to regions of bad curvature, we carry out the $\lambda$ integration in the second term of the denominator. To evaluate the factor $|\overline{\delta \phi}|^{2}$ which is the modulus squared of the bounce-averaged potential, we use the Cauchy-Schwarz inequality

\begin{equation}
    |\overline{\delta \phi}|^{2} \leq \overline{|\delta \phi|^{2}} 
\end{equation}
\citep{Helander2013a} for which the equality holds in the case of a constant potential. In fact, the case of equality is an upper bound on the destabilizing effect of the denominator, so we can use it to estimate the strongest effect of trapped electrons on the mode growth. Assuming a constant potential but this time in the bad curvature region and doing the $\lambda$ integration over the region of finite potential $\delta \phi$, we find the field line average of $\delta \phi$ squared over B with a spatially varying trapped particle fraction $\eps(\ell) = \sqrt{1-B(\ell)/B_{max}}$, where contributions to the integral outside of the bad curvature region are ignored. Note that this is equivalent to using a trial function for the potential which is a step function located entirely in the bad curvature region. The result is:
\begin{align} \label{eqn:tpf}
\omega = \pm \sqrt{-\os_{i} \langle \hat{\o}_{di} \rangle_{\delta \phi
}/ \langle (1-\eps(\ell) \rangle_{\delta \phi} }
\end{align}
where, as in \cite{Costello2025}, the volume average including the mode potential is
\begin{equation}
    \langle (\cdot\cdot\cdot) \rangle_{\delta \phi} = \int_{-\infty}^{\infty} (\cdot\cdot\cdot) |\delta \phi|^{2} \frac{d \ell}{B}.
\end{equation}
For the purposes of optimization, we can approximate this average with the definition $\langle (\cdot\cdot\cdot)\rangle_d$ used in equation (\ref{eqn:galpha}), since this estimates the effect of mode localization in the region of peak bad curvature. In doing so we ignore the factor $1/B$ in the $\ell$ integrals in both the numerator and denominator for simplicity.

Expression (\ref{eqn:tpf}) suggests that we can shift trapped particles to a region of zero or good curvature in order to increase mode inertia (i.e. stabilize the mode), in line with work such as \cite{Proll2016} and \cite{Gerard2023}. An idealized example of this would be the limit of a very narrow square well magnetic field where for the most part $B=B_{max}$ with a small region of $B=B_{min}$, in which the curvature is either zero or stabilizing. Such magnetic fields can be achieved by optimizing quasi-isodynamic configurations to have an ``inverse mirror'' (IM) shape as shown in Fig. \ref{fig:IMB}(c). Modern optimization of quasi-isodynamic designs \citep{Subbotin2006,Sanchez2023,Goodman2024} has tended to lead towards configurations with a broad, flat magnetic field minimum that traps a large number of particles in a region of small field line curvature. The magnetic field around the corresponding minimum of $B$ tends to be relatively straight and wide in real space while the maximum of $B$ tends to be narrow and exhibit bean-shaped cross sections reminiscent of that seen in Wendelstein 7-X configurations. However, the IM magnetic field is another possible solution of the near-axis equations in QI geometry.  While this type of configuration may struggle to fulfill the max-J property, it nonetheless has the tendency to push bad curvature close to the maximum of $B$ such that $\langle \eps(\ell) \rangle \rightarrow 0$. In this limit it may be possible to achieve similar levels of ITG stability without enforcing the maximum J property, as we shall later see. We emphasize that it is not required for a magnetic field to be in this extreme limit to achieve reasonable stabilization, merely that it is one way to stabilize ITG modes via expression (\ref{eqn:tpf}) at finite mirror ratio.

\section{Omnigenity targets}\label{sec:QI}

Here we discuss the general optimization strategy employed in the paper to achieve better ITG stability in QI configurations. We target omnigenity, the property of zero bounce-averaged radial particle drifts, with poloidally closing contours of magnetic field strength. In practice, the resulting configurations are approximately quasi-isodynamic, but with poloidally straight contours near the minimum of the magnetic field. This behavior results because we prescribe fixed toroidal angles for the minimum and maximum of the magnetic field strength in Boozer coordinates, but at locations away from the extrema the field contours are allowed to "wobble" in the Boozer plane with excursions in the poloidal and toroidal directions, in typical QI fashion. On every magnetic field line, using now $\phi_{B}$ as the field-line following coordinate, we ask that the magnetic field achieves its maximum at the locations $\phi_{B}=0$ and $\phi_{B}=2\pi/nfp$, and its minimum at the locations $\phi_{B}=\pi/n_{fp}$, with $n_{fp}$ the number of field periods and $\phi_{B}$ the toroidal Boozer angle \citep{Boozer1981}. We also ask that the magnetic field strength $B$ decreases (increases) monotonically from the field maximum (minimum) to the minimum (maximum) on every field line, penalizing cases where the derivative $B^{\prime}(\l)$ lies below or above a target value, depending on whether the point is in the left or right side of the $B$ well. We target a mirror ratio near the magnetic axis of $25\%$,
\begin{equation}
    f_{mirror} = ((B_{max}-B_{min})/(B_{max}+B_{min}) - 0.25)^{2} \label{eqn:mirror}
\end{equation}
with $B_{max}$ the global maximum and $B_{min}$ the global minimum on the surface, while the target for the maxima and minima is 
\begin{equation}
    f_{\text{extrema}} = \sum_{\alpha} \left[ (B_{\alpha}(\phi_{B}=0)-\text{max}[B_{i}(\phi_{B})])^{2} + (B_{\alpha}(\phi_{B}=\pi/n_{fp})-\text{min}[B_{\alpha}(\phi_{B})])^{2} \right] \label{eqn:extrema}
\end{equation}
where $B_{\alpha}(\phi_{B})$ is the field line dependence of the magnetic field strength for the field line label $\alpha$. Finally, the monotonicity target, which sums over all locations $\phi_{j}$ for the field line label $\alpha$ and evaluates $B_{\alpha,j}=B_{\alpha}(\phi_{j})$, reads 
\begin{equation}
f_{\text{mono}} = \sum_{\alpha,j} \Theta_{\Delta B}((B_{\alpha,j+1}-B_{\alpha,j})-t_{\Delta B})((B_{\alpha,j+1}-B_{\alpha,j}) - t_{\Delta B})^{2} \label{eqn:mono}  
\end{equation}
where to the left of the actual minimum of $B_{\alpha}$ the desired floor on the change in $B$ is $t_{\Delta B}=-|t_{\Delta B,0}|$ and to the right of it, $t_{\Delta B}=|t_{\Delta B,0}|$, with $t_{\Delta B,0}$ usually a small value close to $10^{-3}$ for $128$ points in $\phi$. $\Theta_{\Delta B}$ is the Heaviside step function. This target not only enforces monotonicity but also tries to remove numerically difficult points where $\partial{B}/\partial{\phi_{B}}$ is zero identically.

\subsection{Small Orbit widths}

It has been argued by \cite{Plunk2024} and shown via optimization \citep{Goodman2024} that QI configurations with low neoclassical transport tend to have robust zonal flow suppression of turbulence. Theoretically it was argued that this effect comes from the small orbit width of particles in QI fields (compared to tokamaks or quasisymmetric stellarators), related to the relatively small distance between particle bounce points. By choosing QI configurations as candidates for turbulence optimization, we seek to leverage this property in addition to the CG and growth rate strategies discussed earlier. Having an additional contour at the minimum of $B$ be poloidally straight ({\em i.e.} a contour of constant $\phi_{B}$) in the Boozer plane also seems advantageous from this perspective. This forces the magnetic surface to achieve low geodesic curvature as the poloidal derivative of $B$ is small in the vicinity of $B_{\text{min}}$. Small geodesic curvature has been known to be favorable for zonal flow reduction of electrostatic turbulence \citep{Nunami2013,Nakata2022,Nishimoto2024}, tied to the damping of geodesic acoustic modes (GAMs), and the following results suggest this approach indeed works.


Since targeting the minimum of $B$ to lie at constant toroidal angle seems distinct from previous QI optimizations, we would like to classify it differently. To delineate this case, we define an ``I-number'' (I as in isodynamic) counting the number of poloidally straight contours, which is $I=1$ in the case of a standard QI stellarator (one contour at $B_\text{max}$) and $I = 2$ approximately in the case of the new optimized configurations presented here (an additional contour at $B_\text{min}$). Note that quasi-isodynamicity \citep{Cary1997,Plunk2019} requires poloidally straight contours to come in pairs ($\phi = \phi_\text{min} \pm \phi_\text{str}$) if they are not located at the minimum, with $\phi_{min}$ the location of the minimum of $B$. Omnigenous magnetic fields with no poloidally straight contours such as piecewise omnigenous configurations (Velasco et al.), the various configurations of Wendelstein 7-X, tokamaks, or quasisymmetric stellarators have $I=0$.

\subsection{Radial drift}

The primary omnigenity target minimizes a version of the quantity \epseff \:  [\cite{Nemov1999}] to reduce the net radial drift of particles in accordance with omnigenity. Instead of taking an average over a very long distance of a single field line (in the limit of the distance $s$ going to infinity), we sample a large number of field lines ranging from $\phi=0$ to $\phi=2\pi/n_{fp}$ and assume that over each field period a single well in the magnetic field exists, such that additional bounce wells are not taken into the calculation. This is enforced with targets (\ref{eqn:mono}) and (\ref{eqn:extrema}). The target reads

\begin{align}
    f_{\overline{\eps}_{\text{eff}}} = \sum_{s_{0}} \overline{\eps}_{\text{eff}}^{2}(s_{0}) \label{eqn:epseff} \\
    \overline{\eps}_{\text{eff}}^{3/2} = \pi R_{0}^{2}\left( \sum_{\alpha} \int_{\l_{-}(\alpha)}^{\l_{+}(\alpha)}\frac{d\l}{B} \right)\left( \sum_{\alpha} \int_{\l_{-}(\alpha)}^{\l_{+}(\alpha)} \bnabla \psi \frac{d\l}{B}\right)^{-2}\sum_{\alpha} \int^{B_{max}(\alpha)}_{B_{min}(\alpha)}db^{\prime} \Theta_{B}(b^{\prime}-B) \frac{\hat{H}^{2}}{\hat{I}} \nonumber \\
    \hat{H}=\frac{1}{b^{\prime}}\int_{\l_{-}(\alpha)}^{\l_{+}(\alpha)}\sqrt{b^{\prime}-\frac{B}{B_{0}}}\left(4\frac{B_{0}}{B}-\frac{1}{b^{\prime}}\right)\bnabla \psi \kappa_{g} \frac{d\l}{B} \nonumber \\
    \hat{I} = \int_{\l_{-}(\alpha)}^{\l_{+}(\alpha)}\sqrt{1-\frac{B}{B_{0} b^{\prime}}}\frac{d\l}{B} \nonumber 
\end{align}
with $R_{0}$ a fixed major radius used to stop the optimizer from reducing the aspect ratio from deflating the magnitude of $\epseff$, $b^{\prime}=B/B_{0}$ measures a specific value of B lying between the local $B_{min}$ and $B_{max}$ for a specified field line label $\alpha$, which contains the toroidal interval $\{0,2\pi\}$ in PEST coordinates, and $B_{0}$ is a reference value of $B$. The step function $\Theta_{B}$ is $1$ when $b^{\prime} \geq B$ and zero otherwise.

\section{Optimization}

\begin{figure}
    \centering
    \includegraphics[width=0.38
    \linewidth]{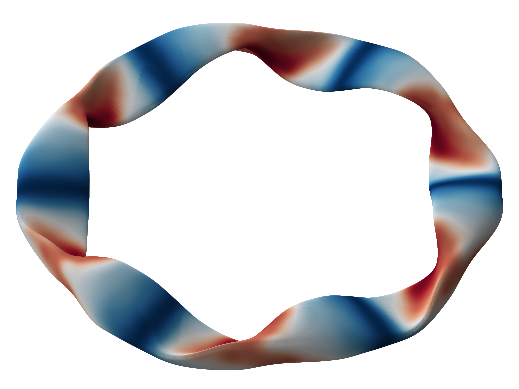}
    \includegraphics[width=0.30\linewidth]{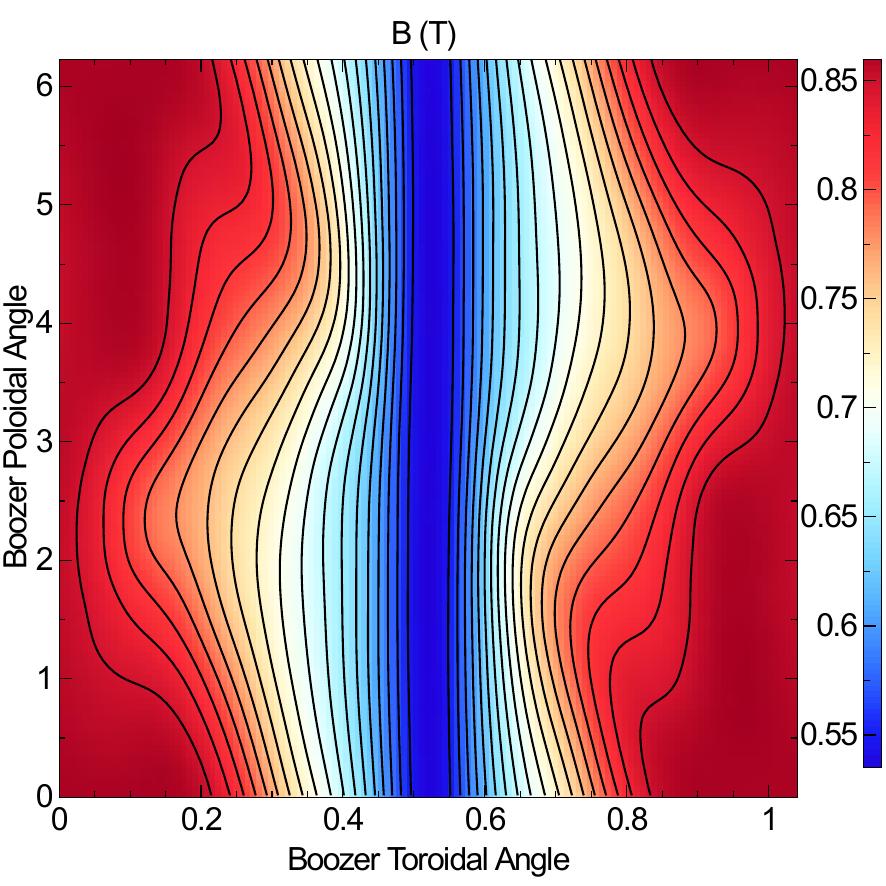}
    \includegraphics[width=0.30\linewidth]{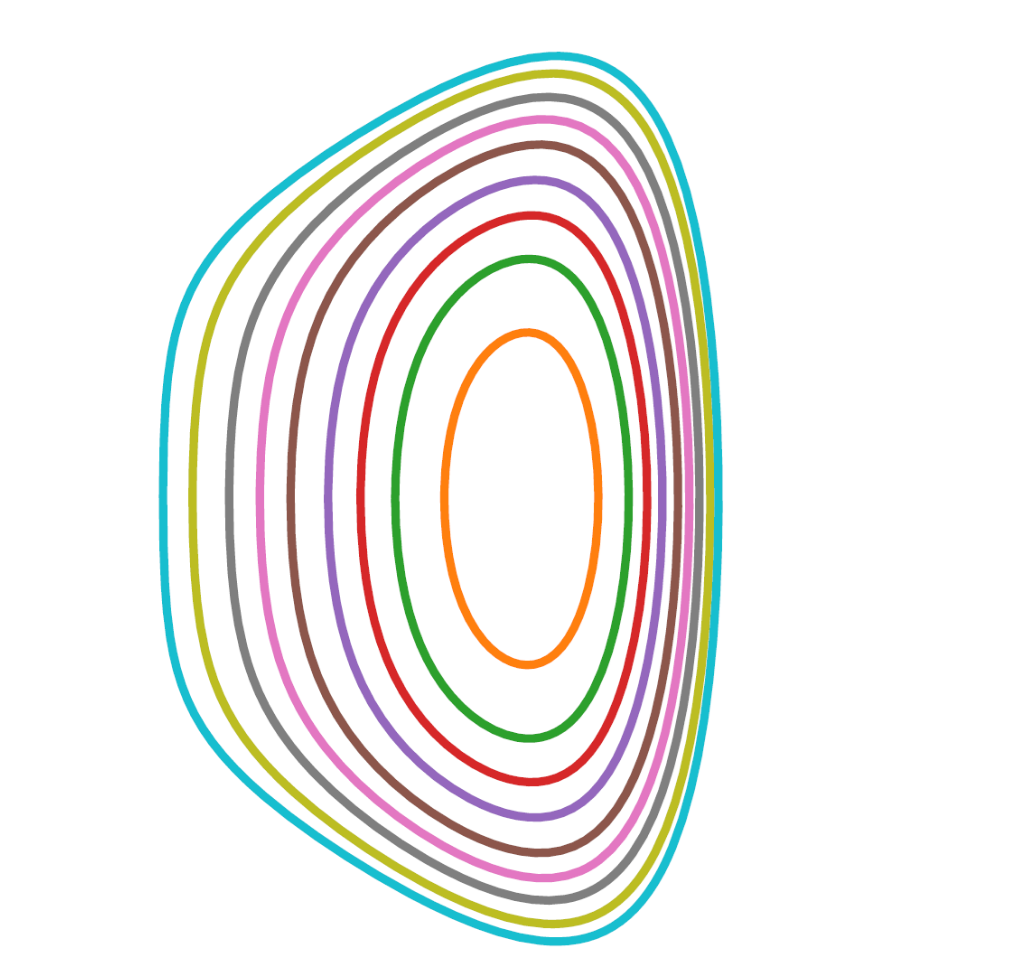}
    \caption{The IM configuration with $a/L_{T,\text{crit}}=2.09$. Left: boundary surface with magnetic field strength in color. Center: Magnetic field line contours and field strength at $r=0.5a$. Right: cross section with several surfaces plotted at zero toroidal VMEC angle, at or near the maximum field strength. Note the higher compression of the surfaces on the outboard where the curvature drive is minimal.}
    \label{fig:CG1.4}
\end{figure}

\begin{figure}
    \centering
    \includegraphics[width=0.60\linewidth]{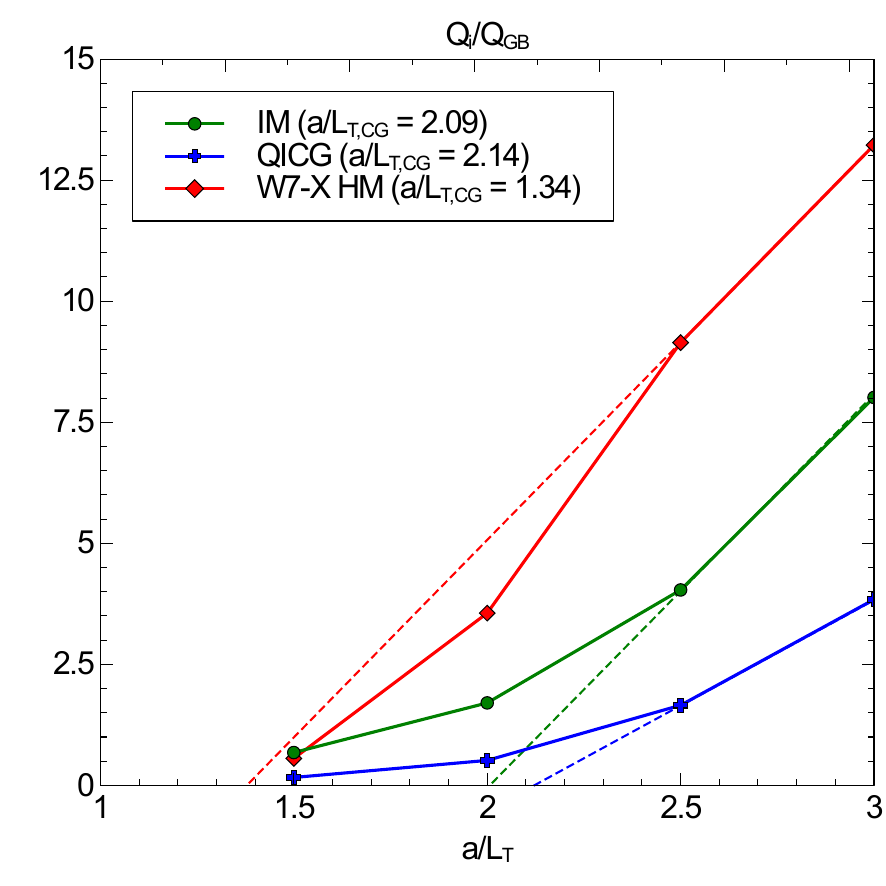}
    \caption{Nonlinear heat flux calculations with adiabatic electrons for the three configurations discussed in the paper, scanning in temperature gradient $a/L_{T}$ at $r/a=0.5$. Two flux tubes (at $\alpha=0$ and $\alpha = \pi/n_{fp}$) were simulated and the more unstable of the two are plotted, depending on the case. For QICG and IM, $\alpha=\pi/n_{fp}$ is more unstable, while for W7-X, $\alpha=0$ is more unstable. A naive extrapolation from the two points at $a/L_{T}=(2.5,3.0)$ down to $Q=0$ yields CGs which are in the neighborhood of the model predictions from Eqn. (\ref{eqn:CGopt}). Notice that the IM configuration has a more pronounced ``foot'' \citep{Zocco2018} below the calculated CG suggesting that more extended ITG modes survive in this geometry.}
    \label{fig:QAdiabatic}
\end{figure}

\begin{figure}
    \centering
    \includegraphics[width=0.99\linewidth]{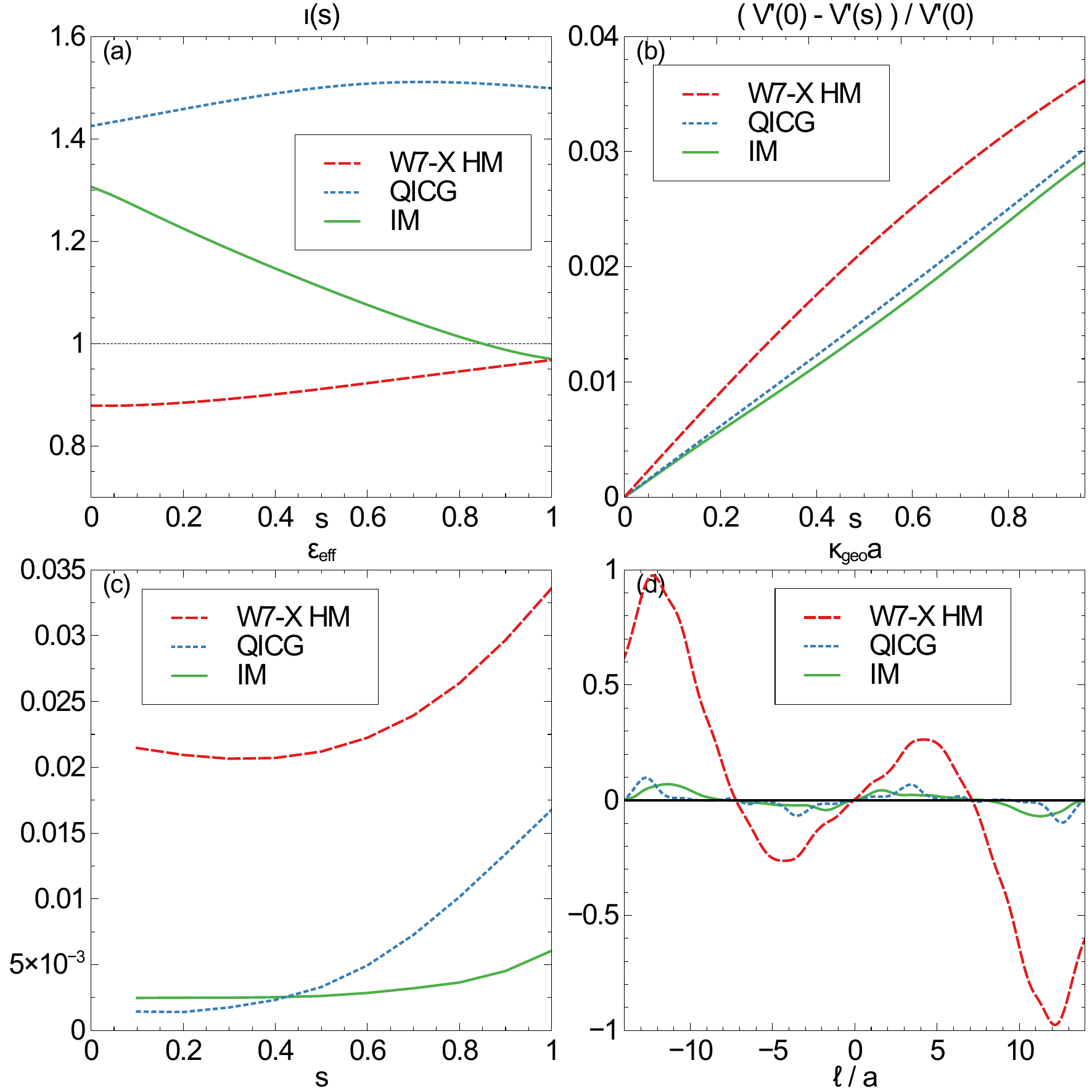}
    \caption{Profiles for the three configurations discussed in the paper at $2\%$ $\beta$. (a) Rotational transform as a function of the radial coordinate $s = (r/a)^2$. (b) Magnetic well ($V$ is the volume enclosed by a flux surface) as a function of $s$. (c) Neoclassical transport coefficient $\epseff$ \citep{Nemov1999} as a function of $s$. (d) Geodesic curvature $\kappa_{\text{geo}}a=a(\mathbf{B} \times \bnabla B) \cdot \bnabla \psi / (B^{2} |\bnabla \psi|) $ at the surface $s=0.25$ and field line $\alpha_{0}=0$.}
    \label{fig:eps}
\end{figure}

Results from using the simsopt optimization framework \citep{Landreman2021} are summarized in Figures \ref{fig:CG1.4}, \ref{fig:QAdiabatic}, and \ref{fig:eps}, comparing two six-field-period optimized QI configurations ($I=2$, with straight minima and maxima of $B$) with the high mirror W7-X (HM) configuration ($I=0$, with no poloidally straight contours). Fig. \ref{fig:QAdiabatic} shows the results of nonlinear flux tube gyrokinetic simulations using the GENE code \citep{Jenko2000a}, scanning in temperature gradient, while Fig. \ref{fig:eps} shows various relevant quantities, such as the rotational transform, vacuum magnetic well, and neoclassical transport coefficients, as well as profiles of geodesic curvature along a magnetic field line. Flux tube metrics for gyrokinetic simulations are shown in Fig. \ref{fig:gyyplots}, with discussion of the different behavior for each configuration in the caption. We see in Fig. \ref{fig:QAdiabatic} that a simple fit of the heat flux in each case to a line, using the data at large gradients $a/L_{T}$, can be extrapolated to zero heat flux at locations which are quite close to the model predictions from \ref{eqn:CGopt}. We believe this is because the effects of Landau damping and $\bnabla \alpha$ -- the dominant stabilizing mechanisms near marginality for localized modes -- are more or less accurately modelled by the CG formula \citep{Roberg-Clark2023}. If we then assume that the unstable mode with this CG sets the slope $\partial Q / \partial (a/L_{T})$, the linear fit to $Q$ (see e.g. \cite{Nies2026}) could indicate the threshold for this mode. More generally, the results for the three configurations shown are in line with the general stellarator turbulence pictures discussed in e.g. \citep{Zocco2022} or \citep{Faber2015}, for which heat flux ramps up in a non-linear fashion as $a/L_{T}$ is increased above the absolute CG of the system. Well above the CG, the heat flux resembles the picture of ``stiff'' transport seen in tokamaks, allowing us to attempt extrapolation to the CG.

\subsection{QI configurations}\label{sec:CG2}

The starting point for the two new optimized QI configurations was a five-field-period configuration with zero axis helicity. We try to optimize for a high CG using the aforementioned targets, ignoring for the moment growth rates above the threshold. We find that it's necessary to optimize by checking $32$ flux tubes spanning 2 toroidal field periods to catch the most unstable spot on the flux surface. The optimization is carried out targeting volume-averaged $\beta$ to be $2\%$ using artificial, linear pressure profiles that peak at $r=0$ and are zero at $r=a$. The edge toroidal flux is allowed to vary as an optimization parameter to keep $\beta$ at this value. We note here that both new configurations possess a low value of $\epseff$ as well as small geodesic curvature in comparison to the HM configuration (Fig. \ref{fig:eps}).

\begin{figure}
    \centering
    \includegraphics[width=0.45\linewidth]{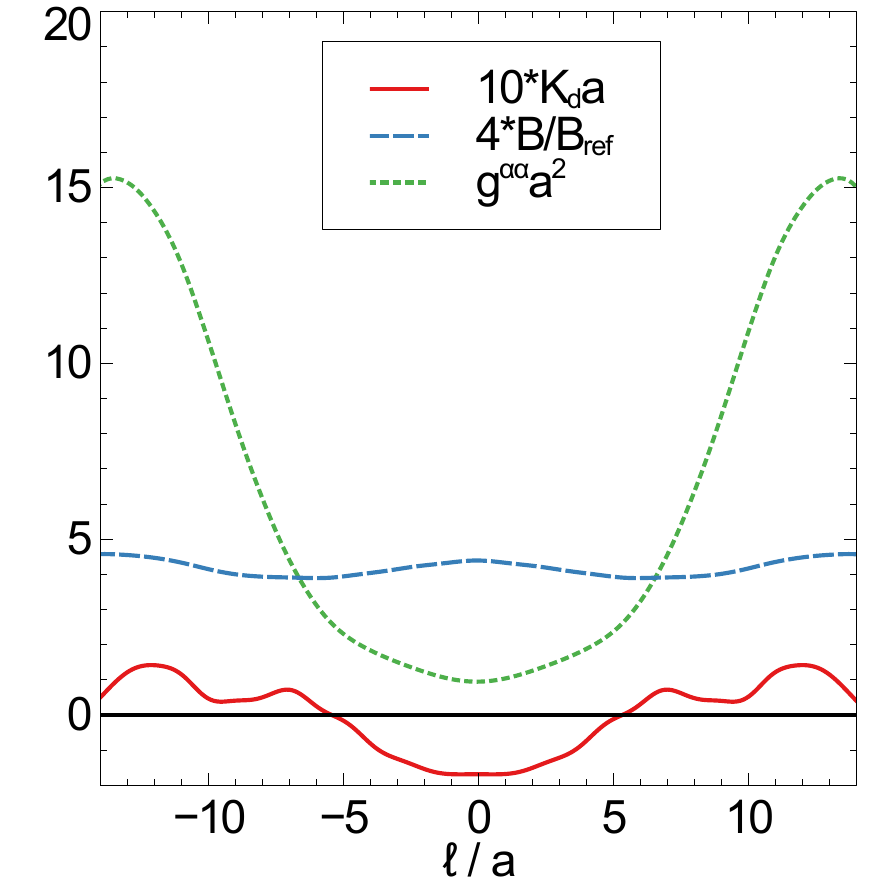}
    \includegraphics[width=0.45\linewidth]{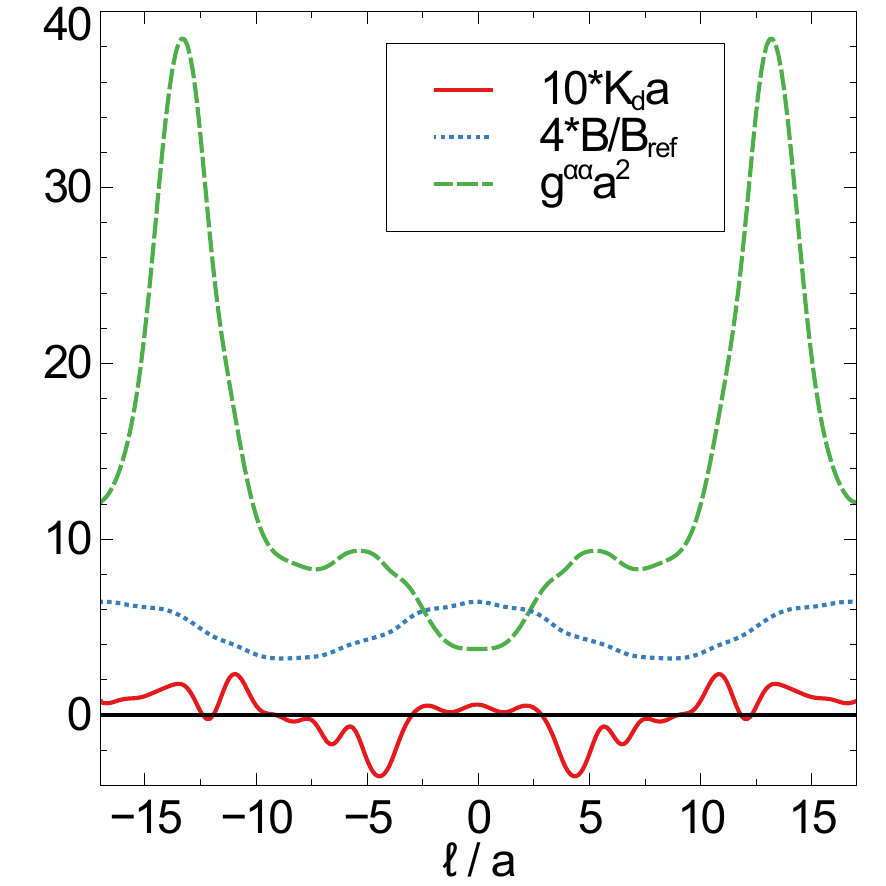}
    \includegraphics[width=0.45\linewidth]{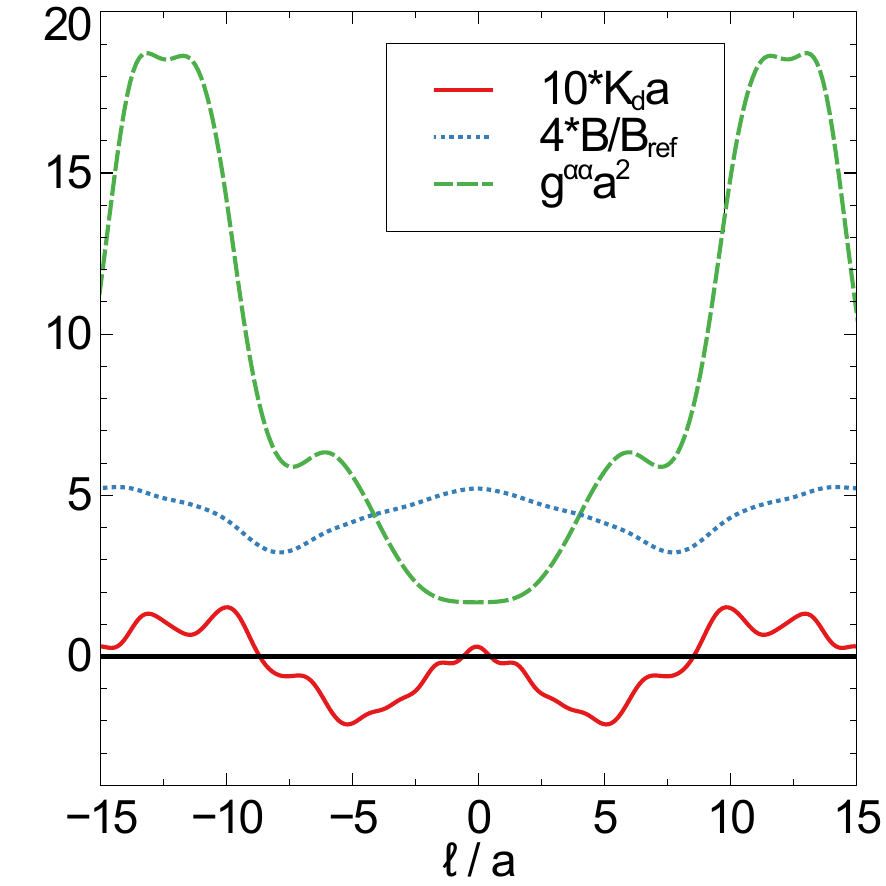}
    \caption{Geometric quantities entering the gyrokinetic equation for the three configurations, for a flux tube centered at the outboard midplane, including drift curvature, normalized magnetic field strength, and $\bnabla \alpha$. Top Left: For W7-X HM, the metrics somewhat resemble those of a quasisymmetric stellarator near the outboard midplane (no curvature gap at $\l=0$). Top Right: The same but for the QICG configuration with $a/L_{T,\text{crit}}=2.14$, for which the ``mode splitting'' is clearly visible in the curvature quantity $K_{d}$, which shows two very separate bad curvature wells near the outboard midplane (near $\ell \sim \pm 4 a$). Bottom Center: The IM configuration, similar to the previous case but with less pronounced splitting. The bad curvature has noticeably shifted away from the minimum of  $B$.}
    \label{fig:gyyplots}
\end{figure}

The first new configuration, which we refer to as QICG, was optimized for QI quality (see previous section), a vacuum magnetic well, an $\iota$ ranging from $1.35$ to $1.50$, and an aspect ratio of $14$. The targets used were:
\begin{align}
    f_{QICG} = f_{well} + f_{iota} + f_{CG} + f_{asp} \nonumber \\ + f_{mono} + f_{extrema} + f_{mirror} + f_{\beta} + f_{\overline{\eps}_{\text{eff}}},
\end{align} 
with \begin{equation}
    f_{asp}(A_{t}) = (A - 14)^{2},
\end{equation}  ($A$ is the aspect ratio), 
\begin{equation}
    f_{\iota}=\left[(\iota(s_{0}=0.0) - 1.35)^{2} + (\iota(s_{0}=1.0) - 1.49)^{2} \right],
\end{equation}
and 
\begin{equation}
    f_{well}=\sum_{r}\left[(1+V^{\prime\prime}(r)/0.01)\Theta(1+V^{\prime\prime}(r)/0.01) \right]^{2},
\end{equation}
as in \cite{Roberg-Clark2023} with $r/a=\{0.0,0.25,0.95\}$, and 
\begin{equation}
    f_{\beta}=(\beta - 0.02)^{2},
\end{equation}
with $\beta$ the volume-averaged value calculated from VMEC. The targeted surfaces $r/a$ for $\eps_{\text{eff}}$ were $r/a=(0.02,0.10,0.20,0.30,0.40,0.50,0.60,0.70,0.80,0.90)$ and varying weights were used for all of the terms over the course of the optimization to prioritize at first QI quality and then the CG metric (\ref{eqn:CGf}). The vacuum magnetic well (after removing pressure from the equilibrium) is roughly 0.25 percent (not shown), while at $2\%$ $\beta$ it has a low $\epseff$ of roughly $0.2\%$ on axis. Fig. \ref{fig:QAdiabatic} suggests that transport is relatively low below $a/L_{T}=2.14$ and stays moderate above that point.  The primary effect leading to the large CG of QICG is the mode splitting effect, which is clearly demonstrated in the top right-hand plot of Fig. \ref{fig:gyyplots}, with an additional boost to $g^{\alpha \alpha}$ through a localized shear effect.

\begin{figure}
    \centering
    \includegraphics[width=0.6\linewidth]{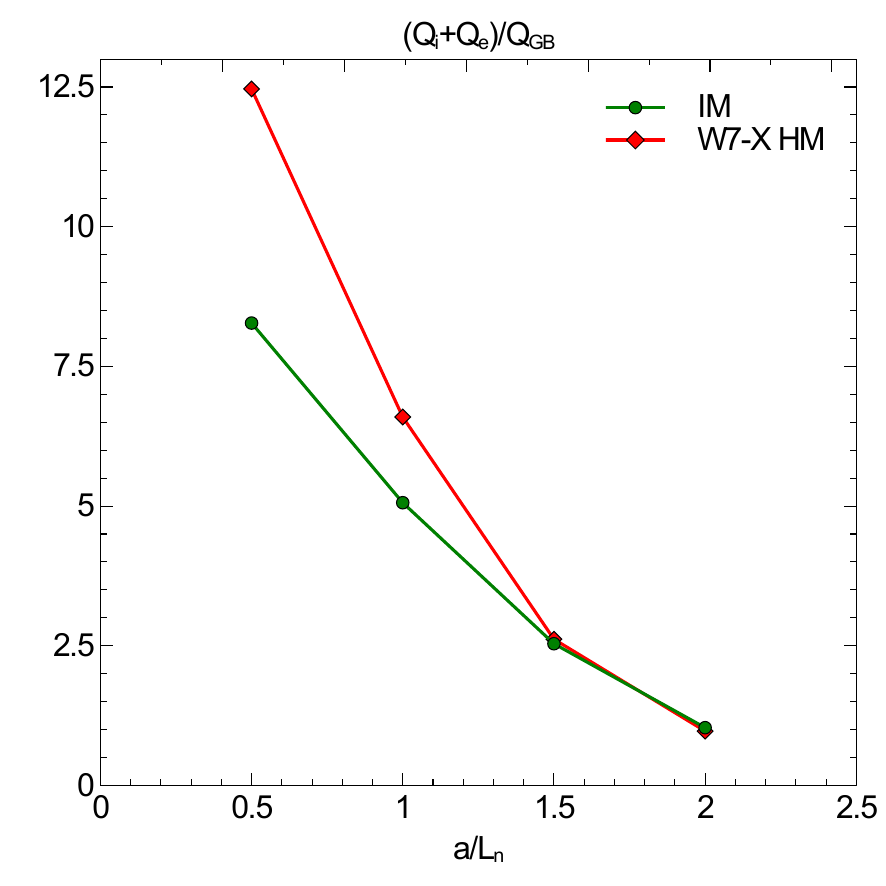}
    \caption{Nonlinear flux tube ITG simulations with kinetic electrons, a fixed temperature gradient $a/L_{T}=a/L_{T,i}=a/L_{T,e}=3$ and variable density gradient $a/L_{n}$ for both species at half radius at the most unstable location. For W7-X, the flux tube is $\alpha=0$ and for IM, it is $\alpha=\pi/n_{fp}$ (see caption for Fig. \ref{fig:QAdiabatic}). The heat flux for the IM case remains below or comparable to that of the W7-X high mirror configuration up to $a/L_{n}=2$.}
    \label{fig:itgcg1.4}
\end{figure}

Since we have not taken into account stability above the threshold, initial nonlinear results (not shown) reveal prohibitively large heat fluxes, likely exacerbated by the mode inertia effect. We re-optimize the QICG configuration so that the ITG growth rates are not too large, yielding the configuration we call ``IM''. To reduce linear ITG growth growth rates with kinetic electrons, we construct a simple target that combines the factors discussed in Sections \ref{sec:stiff} and \ref{sec:kinetic}, which reads
\begin{equation} \label{eqn:stiff}
    f_{Q} = (\left [ \langle |\bnabla \alpha| \rangle (1 - \langle \eps(\ell) \rangle_d ) \right]^{-1} - t_{Q})^{2},
\end{equation}
with the averages calculated as in expression (\ref{eqn:galpha}). Using a new set of targets
\begin{equation}
    f_{IM} = f_{QICG} + f_{Q},
\end{equation}
with $t_{Q}=2.0$ (QICG values exceeded 3) as well as a CG of 2.0 (eqn. \ref{eqn:CGopt}), we arrive at the configuration in Fig. \ref{fig:CG1.4}, with a significant positive (tokamak-like) shear, a rotational transform ranging from 1.3 to 0.98 (near targeted values $1.25$ to $1.01$), an $\epseff$ of roughly $0.2\%$ on axis (\ref{fig:eps}), $A=13$ (as targeted) and a vacuum magnetic well of $0.17\%$. In particular, the positive shear seems to be helpful for the mode inertia effect, as the bad curvature region appears to be shifted away from the minimum of $B$. We see that the configuration has favorable heat fluxes compared to W7-X HM when kinetic electrons are included [Fig. \ref{fig:itgcg1.4}]. 

\section{Conclusion} \label{sec:conc}

This paper has presented a series of physics-motivated model targets for turbulence optimization in stellarators as well as a general strategy for increasing critical gradients that we refer to as mode splitting. That the targets guide the optimization towards ITG stability is shown with a series of example physics configurations that also possess or nearly possess a vacuum magnetic well. The first configuration shown has an approximate CG of $a/L_{T}=2.14$, an unusually high value. The second configuration, which resembles an inverse mirror, maintains a similar CG while sacrificing some nonlinear stability for the case of adiabatic electrons, yet performs well in the presence of kinetic electrons via suppressing the so-called mode inertia effect \citep{Costello2025}. Another property that distinguishes these configurations from previous QI results is that the minimum of the magnetic field strength is approximately poloidally straight in the Boozer plane. We argue that this is helpful in that deeply trapped particles will experience low values of geodesic curvature in their orbits. It remains to be seen how well $\alpha$-particle confinement can be improved in such configurations, which will require more careful attention towards improving QI quality, beyond simply targeting $\epseff$. We note here that analytic progress using optimal mode analysis \citep{Plunk2023,Costello2025a} and model geometries \citep{Rodriguez2025} may also be helpful in refining critical gradient estimates in the future.

\textit{Acknowledgments--} 
We would like to thank A. Goodman for helpful tips on optimizing QI configurations, in vacuum and at finite $\beta$, and calculating vacuum magnetic wells in simsopt, as well as sharing VMEC equilibrium plotting routines. We thank P. Helander, P. Costello, M. Landreman, W. Sengupta, E. Paul, E. Rodr\'iguez, and R. Nies for helpful advice and discussions.

\textit{Conflict of interest--} The authors report no conflict of interest.

\textit{Funding--}This work has been carried out within the framework of the EUROfusion Consortium, funded by the European Union via the Euratom Research and Training Programme (Grant Agreement No 101052200 — EUROfusion). Views and opinions expressed are however those of the author(s) only and do not necessarily reflect those of the European Union or the European Commission. Neither the European Union nor the European Commission can be held responsible for them.

\bibliographystyle{jpp}
\bibliography{library.bib}

\end{document}